\UseRawInputEncoding
\documentclass[superscriptaddress, notitlepage, reprint,twocolumn]{revtex4-1}
\usepackage[english]{babel}
\usepackage{amsmath,amsthm}
\usepackage{amsfonts}
\usepackage[pdfborder={0 0 0}, colorlinks=true, urlcolor=blue, linkcolor=blue, citecolor=blue]{hyperref}
\usepackage{color}
\usepackage{graphicx}
\usepackage{float}
\usepackage{subfigure}
\usepackage{lipsum}
\usepackage{epstopdf}
\usepackage{dcolumn}
\begin{document}

\title{Quantum metrology in a driven-dissipation down-conversion system beyond the parametric approximation}
\author{Dong  Xie}
\email{xiedong@mail.ustc.edu.cn}
\author{Chunling Xu}
\email{xuchunling@guat.edu.cn}
\affiliation{College of Science, Guilin University of Aerospace Technology, Guilin, Guangxi 541004, People's Republic of China}

\begin{abstract}
We investigate quantum metrology in a degenerate down-conversion system composed of a pump mode and two degenerate signal modes. In the conventional parametric approximation, the pump mode is assumed to be constant, not a quantum operator. We obtain the measurement precision of the coupling strength between the pump mode and two degenerate signal modes beyond the parametric approximation. Without a dissipation, the super-Heisenberg limit can be obtained when the initial state is the direct product of classical state and quantum state. This does not require the use of entanglement resources which are not easy to prepare. When the pump mode suffers from a single-photon dissipation, the measurement uncertainty of the coupling strength is close to 0 as
the coupling strength approaches 0 with a coherent driving. The direct photon detection is proved to be the optimal measurement. This result has not been changed when the signal modes suffer from the two-photon dissipation. When the signal modes also suffer from the single-mode dissipation, the information of the coupling strength can still be
obtained in the steady state. In addition, the measurement uncertainty of the coupling strength can also be close to 0 and become independent of noise temperature as the critical point between the normal and superradiance phase approaches. Finally, we show that a driven-dissipation down-conversion system can be used as a precise quantum sensor to measure the driving strength.
\end{abstract}
\maketitle

\section{Introduction}
Quantum metrology mainly studies how to use quantum resources to improve the precision of parameter measurement over classical resources\cite{lab1,lab2,lab3,lab4,lab5,lab6}.
The quantum resources, such as superposition and entanglement, offer a
possibility to make the measurement precision with linear generator scale as $1/N$ with $N$ being the total average number of photons, which is the Heisenberg scaling\cite{lab7}. However, classical resources only get the scaling $1/\sqrt{N}$ at most, which is the standard quantum limit\cite{lab4}. By using nonlinear generator or time-dependent evolutions, the super-Heisenberg limit can be obtained\cite{lab7a1,lab7a2,lab7a3,lab7a4}.

The parametric down-conversion is relatively easy to become a source of entangled photon pairs\cite{lab8}, which plays an important role in quantum information processing.
Moreover, the parametric down-conversion can be used to build optical parametric amplifiers, which is the main difference between nonlinear SU(1,1)-interferometers and the linear SU(2)-interferometers.  SU(1,1)-interferometers offer a platform for multi-photon absorption\cite{lab9,lab10,lab11,lab12}, phase measurement\cite{lab12a}, loss-tolerant quantum metrology\cite{lab13,lab14}, Wigner function tomography\cite{lab15}, and quantum state engineering\cite{lab16,lab17}.

Most of the previous work has used the parametric approximation\cite{lab18} to deal with the down-conversion process, where the number of pump mode is assumed to be constant. This parametric approximation can give an exact description of the system dynamics at short times, but the resulting state will become invalid as the number of pump photons decreases with time. Recently, K. Chinni and N. Quesada\cite{lab19} used the cumulant
expansion method, perturbation theory, and the full numerical simulation of systems to deal with the closed down-conversion process beyond the parametric approximation. However, quantum metrology beyond parametric approximation in parametric down-conversion systems, especially in the case of driven-dissipation, has not been studied at present.

This inspires us to try to fill the gap. In this article, we investigate quantum metrology in a degenerate down-conversion system composed of a pump mode and two degenerate signal modes. In the closed system, we analytically obtain the quantum Fisher information of  the nonlinear coupling strength $g$ between the pump mode and the signal modes. It shows that the super-Heisenberg limit can be obtained without using preparative entangled or squeezed states,  just the signal modes in the Fock state.
The reason for obtaining the super-Heisenberg  limit is the nonlinear generator of $g$.
When the pump mode suffers from a single-photon dissipation,
the measurement uncertainty of the coupling strength is close to 0 as the coupling strength approaches 0 due to a continuous coherent driving. By comparing with the results achieved by the quantum Fisher information, the direct photon detection is proved to
be the optimal measurement. This result has not been changed when the signal modes suffer from the two-photon dissipation. When the signal modes also suffer from the single-mode dissipation, the information of very weak coupling strength can still be
obtained with a finite precision under steady state. Finally,  the
measurement uncertainty of the coupling strength can also be close to 0 and become independent of the temperature of the noise as the critical point between the normal and superradiance phase approaches.

This article is organized as follows. In section II, quantum metrology in a closed degenerate down-conversion system is studied. In section III, quantum metrology in a driven-dissipation degenerate down-conversion system is studied in different cases. Quantum sensor of the driving strength is explored in section IV.  Finally, we make a simple conclusion and outlook in section V.
\section{degenerate down-conversion system}
We consider that the Hamiltonian describing the
degenerate down-conversion process on resonance ($\omega_1=2\omega_2$)is given by

\begin{align}
H=\omega_1a^\dagger a +\omega_2b^\dagger b+g a b^{2\dagger}+g a^\dagger b^2,
\end{align}
where $a\ (a^\dagger)$ is the annihilation (creation) bosonic operator for
the pump mode with the frequency $\omega_1$, and $b\ (b^\dagger)$ is the annihilation (creation) operator for
the signal mode with the frequency $\omega_2$. The operators $a$ and $b$ satisfy the canonic commutation relation, i.e.,  $[c,c^\dagger]=1$ with $c=\{a,b\}$.  Without loss of generality, the coupling strength $g$ is chosen to be the real quantity and $\hbar=1$ through this article.

In the interaction picture, the Hamiltonian is rewritten as
\begin{align}
\mathcal{H}=g a b^{2\dagger}+g a^\dagger b^2.
\end{align}
For the pure state $|\psi(g)\rangle$, the quantum Fisher information for the parameter $g$ can be calculated by\cite{lab20,lab21}
\begin{align}
\mathcal{F}(g)=4(\langle \partial_\theta\psi(g)|\partial_\theta\psi(g)\rangle-|\langle \psi(g)|\partial_g\psi(g)\rangle|^2),
\end{align}
where $|\psi(g)\rangle=e^{-i g t\mathcal{H}}|\psi(t=0)\rangle$ and $\partial_g=\frac{\partial}{\partial g}$.

 When the initial state is a semi-classical state (direct product of classical state and quantum state) $|\psi(t=0)\rangle=|\alpha\rangle|n\rangle$ with the coherent state $|\alpha\rangle$ and the Fock state $|n\rangle$,
we can obtain that $\mathcal{F}(g)=4[\alpha^2(2n^2+2n+2)+n(n-1)]t^2$. Given by the fixed total number of photons, $\alpha^2+n=N$, the optimal estimation precision $\mathcal{F}_o(g)$ is obtained with $n=2/3N$ for $n\gg1$,
\begin{align}
\mathcal{F}_o(g)=\frac{32}{27} N^3 t^2.
\end{align}

When the initial state is a complete quantum state (direct product of quantum state and quantum state)  $|\psi(t=0)\rangle=|n_1\rangle|n_2\rangle$ with $n_1+n_2=N$, $\mathcal{F}(g)=4[n_1(2n_2^2+2n_2+2)+n_2(n_2-1)]t^2$. The optimal estimation precision $\mathcal{F'}_o(g)$ can be achieved when $n_2=2/3N$ for $n_2\gg1$,
\begin{align}
\mathcal{F'}_o(g)=\frac{32}{27} N^3 t^2.
\end{align}
Obviously, the optimal measurement precision is the same for the complete quantum state and the semi-classical state, i.e., $\mathcal{F}_o(g)=\mathcal{F'}_o(g)$.

For the complete classical initial state $|\psi(0)\rangle=|\alpha_1\rangle|\alpha_2\rangle$, the according quantum Fisher information is $\mathcal{F}(g)=\alpha_2^4 t^2$. Given by the constraint condition $\alpha_1^2+\alpha_2^2=N$, the maximal quantum Fisher information is achieved $\mathcal{F}(g)=N^2 t^2$ when $\alpha_1=0$. It shows that the complete classical state can not perform better than the complete quantum state and the semi-classical state. In the down-conversion system, the super-Heisenberg limit can be obtained only if the initial state of the signal mode is quantum state, where entangled or squeezed states that are difficult to prepare are not required. The reason for obtaining the super-Heisenberg  limit is due to the fact that the generator $ a b^{2\dagger}+ a^\dagger b^2$ is nonlinear.

\section{driven-dissipation system}
\begin{figure}[h]
\includegraphics[scale=0.5]{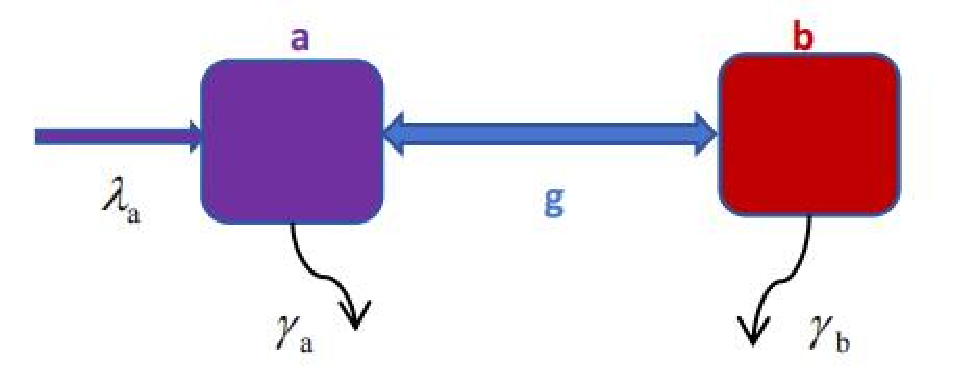}
 \caption{\label{fig.1}Schematic diagram of a dissipative down-conversion system composed of the subsystem $a$ (pump mode) with the frequency $\omega_1$ and the subsystem $b$ (signal mode) with the frequency $\omega_2$. $\lambda_a$ represents the coherent driving strength. $g$ denotes the nonlinear coupling strength between the subsystem $a$ and the subsystem $b$. $\gamma_a$ and $\gamma_b$ are the dissipation rate of the subsystem $a$ and the subsystem $b$, respectively.}
\end{figure}
In this section, we consider that the degenerate down-conversion system suffers from an inevitably dissipative environment, as shown in Fig.~\ref{fig.1}. In order to resist the dissipation, we consider that there is an extra coherent driving $H_d=i\lambda_a(a^\dagger-a)$. The evolution of the density matrix $\rho$ of the driven-dissipation system is described by the master equation
\begin{align}
\dot{\rho}=-i[H+H_d,\rho]+\gamma_a\mathcal{D}[a]\rho+\gamma_b\mathcal{D}[b]\rho,
\label{eq:6}
\end{align}
where the superoperator $\mathcal{D}[c]\rho=\gamma_c[2c\rho c^\dagger-c^\dagger c\rho-\rho c^\dagger c]$ with $c=\{a,\ b\}$.
The corresponding quantum Langevin-Heisenberg equation of a operator $O$ is derived by\cite{lab22,lab23,lab24}
\begin{align}
\dot{O}=i[H+H_d,O]-\sum_{c=a,b}\{[O,c^\dagger](\gamma_cc-\sqrt{2\gamma_c}c_{in})\nonumber\\
-(\gamma_cc-\sqrt{2\gamma_c}c_{in})[O,c]\}.
\label{eq:7}
\end{align}
Utilizing the above equation, let $O=a, b$, we can obtain the evolution dynamics
\begin{align}
\dot{a}&=-ig b^2-\gamma_a a+\sqrt{2\gamma_a }a_{in}+\lambda_a,\label{eq:8}\\
\dot{b}&=-2ig ab^\dagger-\gamma_b b+\sqrt{2\gamma_b }b_{in},\label{eq:9}
\end{align}
where the expected values of the noise operators are $\langle c_{\textmd{in}}(t) \rangle=\langle c_{\textmd{in}}^\dagger(t) \rangle=0$, $\langle c^\dagger_{\textmd{in}}(t)c_{\textmd{in}} (t')\rangle=0$ and $\langle c_{\textmd{in}}(t)c^\dagger_{\textmd{in}} (t')\rangle=\delta(t-t')$ with $c_{\textmd{in}}=\{a_{\textmd{in}},\ b_{\textmd{in}}\}$.

\subsection{In the case of $\gamma_a\gg\gamma_b$}
When the decay rate $\gamma_a$ is much lager than $\gamma_b$ ($\gamma_a\gg\gamma_b$), the pump mode $a$ can be adiabatically eliminated. Let $\dot{a}=0$, we obtain
\begin{align}
 a=(-ig b^2+\sqrt{2\gamma_a }a_{in}+\lambda_a)/\gamma_a.
\end{align}
Substituting the above equation into Eq.~(\ref{eq:7}), we obtain that
\begin{align}
\dot{b}=&-2g^2 b^\dagger b^2/\gamma_a-2ig\lambda_ab^\dagger/\gamma_a -2ig\sqrt{2/\gamma_b }b^\dagger a_{in}\nonumber\\
&-\gamma_b b+\sqrt{2\gamma_b }b_{in}.
\end{align}

By combining Eq.~(\ref{eq:6}) with Eq.~(\ref{eq:7}), we can get the reduced master equation of the signal mode, which is described as
\begin{align}
\dot{\rho}=-i[H_b,\rho]+\gamma_b\mathcal{D}[b]\rho+\kappa\mathcal{D}[b^2]\rho,\label{eq:12}
\end{align}
where $H_b=\frac{g\lambda_a}{\gamma_a}(b^2+b^{2\dagger})$ and $\kappa=2g^2/\gamma_a$.

The steady state can be analytically derived by a complex P-representation solution\cite{lab25}  or  Keldysh-Heisenberg equation\cite{lab26}. The general form can be expressed as
\begin{align}
\langle b^{\dagger l}b^k\rangle=\frac{1}{\mathcal{N}\sqrt{2^{l+k}}}\sum_{m=0}^\infty \frac{1}{m!} F^*_{m+l}F^*_{m+k},\label{eq:13}
\end{align}
where the normalization factor $\mathcal{N}=\sum_{m=0}^\infty \frac{|\mu|^{2m}}{m!} |_2F_1[-m,y;z;2] |^2$, and the function $F_{m+k}=(-\mu)^{m+k}{} _2F_1[-(m+k),y;z;2] $, in which $\mu=i\sqrt{2i\lambda_a/g}$, $y=\gamma_a\gamma_b/4g^2$, and $z=\gamma_a\gamma_b/2g^2$.
The Gauss hypergeometric function ${} _2F_1[-m,y;z;2]$ is described by
\begin{align}
{} _2F_1[-m,y;z;2]=\sum_{n=0}^\infty \frac{(-m)_n(y)_n 2^n}{(z)_nn!},
\end{align}
where the function $(r)_n=\Gamma(r+n)/\Gamma(r)$ and $\Gamma(r)$ denotes the Gamma function.

\subsubsection{In the case of $\gamma_b=0$}
When the decay rate $\gamma_b=0$, the expected value in Eq.~(\ref{eq:13}) can be simplified as
\begin{align}
\langle b^{\dagger l}b^k\rangle=(i\sqrt{\frac{-i\lambda_a}{g}})^l(-i\sqrt{\frac{i\lambda_a}{g}})^k.\label{eq:15}
\end{align}
With the specific measurement operator $M$, the measurement uncertainty of the parameter $g$ can be calculated by the error propagation formula
\begin{align}
\delta^2g=\frac{\langle M^2\rangle-\langle M\rangle^2}{|\partial\langle M\rangle/\partial g|^2}.
\end{align}
For the direct photon detection with $M_d=b^\dagger b$, the uncertainty of the parameter $g$ is derived by the above equation
\begin{align}
\delta^2g=g^3/\lambda_a.\label{eq:17}
\end{align}
For the homodyne detection with $M_h=be^{-i\varphi}+b^\dagger e^{i\varphi}$, the uncertainty of the parameter $g$ is given by
\begin{align}
\delta^2g=2g^3/[\lambda_a(\cos \varphi-\sin \varphi)^2],
\end{align}
When $\varphi=0$, i.e., the Homodyne detection $M_h=b+b^\dagger $, the above measurement uncertainty is optimized,
\begin{align}
\delta^2g=2g^3/\lambda_a.\label{eq:19}
\end{align}
Comparing Eq.~(\ref{eq:17}) with Eq.~(\ref{eq:19}), we find that the direct photon detection performs better than the homodyne detection.
All results show that the uncertainty $\delta^2g$ is close to 0 as the parameter $g\rightarrow 0$. It is due to the fact that the subsystem of the signal mode can not arrive at the steady state when $g=0$. At this point, the total number of photons tends to infinity as $g$ goes to 0, i.e., $N_b=\langle b^\dagger b\rangle=\lambda_a/g\rightarrow \infty$. It is due to the fact that the infinite encoding time makes the measurement uncertainty become 0.
Using $N_b$, Eq.~(\ref{eq:17}) can be reexpressed as
\begin{align}
\delta^2g=\lambda_a^2/N_b^3.
\end{align}
It shows that the super-Heisenberg scaling $1/N_b^3$ of the single photon dissipation of the pump mode can be obtained by local measurement of the signal mode without considering the time consumption.
\subsubsection{Extra two-photon dissipation}
Then, we consider that the signal mode suffers from an extra two-photon dissipation  with dissipation rate $\kappa_e$. The total evolution dynamics of the signal mode is described by
 \begin{align}
\dot{\rho}=-i[H_b,\rho]+\gamma_b\mathcal{D}[b]\rho+(\kappa+\kappa_e)\mathcal{D}[b^2]\rho,
\label{eq:21}
\end{align}

When the decay rate $\gamma_b=0$, the expectated values can be simplified as
\begin{align}
\langle b^{\dagger l}b^k\rangle=(i\sqrt{\frac{-2ig\lambda_a}{\gamma_a(\kappa+\kappa_e)}})^l(-i\sqrt{\frac{2ig\lambda_a}{\gamma_a(\kappa+\kappa_e)}})^k.
\label{eq:22}
\end{align}
Using the direct photon detection, the measurement uncertainty of the parameter $g$ is given by
\begin{align}
\delta^2g=\frac{g(\kappa_e\gamma_a+2g^2)^3}{2\lambda_a(\kappa_e\gamma_a-2g^2)^2}.\label{eq:23}
\end{align}

For the Gaussian state, the quantum Fisher information is derived by\cite{lab27,lab28}
\begin{align}
\mathcal{F}(g)=&\frac{2d^2}{4d^2+1}\textmd{Tr}[(\mathcal{C}^{-1}\partial_g\mathcal{C})^2]+
\frac{8(\partial_g d)^2}{16d^4-1}\nonumber\\
&+\langle \partial_g\mathbf{X}^\top\rangle \mathcal{C}^{-1} \langle \partial_g\mathbf{X}\rangle,
\end{align}
where $\mathbf{X}^\top=(q,p)$ with quadrature operators defined as: $p=\frac{1}{\sqrt{2}}(b+b^\dagger)$, and $q=\frac{1}{i\sqrt{2}}(b-b^\dagger)$. And the entries of the covariance matrix are defined as $\mathcal{C}_{ij}=\frac{1}{2}\langle \mathbf{X}_i\mathbf{X}_j+\mathbf{X}_j\mathbf{X}_i\rangle-\langle {\mathbf{X}_i\rangle\langle\mathbf{X}_j}\rangle$. $d$ is given by $d=\sqrt{\textmd{Det}\mathcal{C}}$.

Based on the expected values in Eq.~(\ref{eq:22}), we obtain that
\[
\mathcal{C}=\frac{1}{2}\left(
\begin{array}{ll}
1\ \ \ 0\\
0\ \ \ 1
  \end{array}
\right ).
\]
Then, we derive that $d=1$. As a result, the quantum Fisher information can be obtained
\begin{align}
\mathcal{F}(g)=\langle \partial_g\mathbf{X}^\top\rangle \mathcal{C}^{-1} \langle \partial_g\mathbf{X}\rangle\\
=\frac{2\lambda_a(\kappa_e\gamma_a-2g^2)^2}{g(\kappa_e\gamma_a+2g^2)^3}.
\end{align}
According to the quantum Cram\'{e}r-rao bound\cite{lab29,lab30,lab31}, the
measurement uncertainty is given by
\begin{align}
\delta^2g\geq 1/\mathcal{F}(g)=\frac{g(\kappa_e\gamma_a+2g^2)^3}{2\lambda_a(\kappa_e\gamma_a-2g^2)^2}.
\end{align}
Comparing with the result obtained by the direct detection in Eq.~(\ref{eq:23}), it obviously shows that the direct detection is the optimal measurement, which can obtain the optimal measurement precision.

The above result shows that the measurement uncertainty is also close  to 0 as $g$ goes to 0. In order to explain this, we calculate the characteristic time for the system to reach steady state.
Let $b=\langle b\rangle+\delta b$, keeping only the linear terms, the evolution of $\langle\delta b\rangle$ is reduced to
\begin{align}
\langle\dot{{\delta b}}\rangle= - \frac{8g \lambda_a}{\gamma_a(\kappa+\kappa_e)}\langle\delta b\rangle.
\end{align}
According to the above equation, the characteristic time to steady state is given by
\begin{align}
\tau=\frac{\gamma_a(\kappa+\kappa_e)}{8g \lambda_a}.
\end{align}
This result shows that the characteristic time $\tau$ is divergent at $g=0$.
The two-photon dissipation does not allow the signal mode to reach the steady state.
The intuitive explanation is that there are subspaces that are not subject to two-photon dissipation.

\subsubsection{Three-level approximation in the case of $\gamma_a\gg\gamma_b\neq0$}
Expanding on the Fock basis $\{|n\rangle, n = 0, 1, 2, . . . \}$, Eq.~(\ref{eq:21}) can be rewritten as
\begin{align}
\dot{\rho}_{n,n'}=&-i\lambda_a g/\gamma_a[\sqrt{(n+1)(n+2)}\rho_{n+2,n'}\nonumber\\
&+\sqrt{(n-1)n}\rho_{n-2,n'}-\sqrt{(n-1)n}\rho_{n,n'-2}\nonumber\\
&-\sqrt{(n'+1)(n'+2)}\rho_{n,n'+2}]+\nonumber\\
&+\gamma_b[\sqrt{(n+1)(n'+1)}\rho_{n+1,n'+1}-\frac{n+n'}{2}\rho_{n,n'}]\nonumber\\
&+\kappa [\sqrt{(n+2)(n+1)(n'+2)(n'+1)}\rho_{n+2,n'+2}\nonumber\\
&-\frac{n(n-1)+n'(n'-1)}{2}\rho_{n,n'}].\label{eq:30}
\end{align}

\begin{figure}[h]
\includegraphics[scale=0.9]{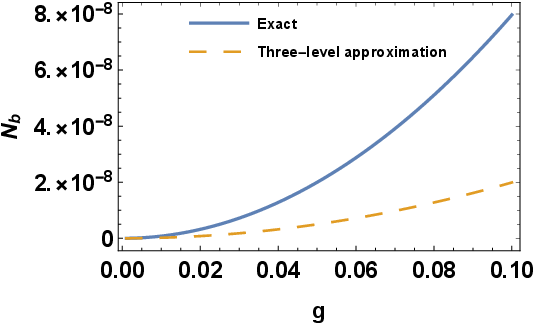}
 \caption{\label{fig.2}Comparison diagram of the expected number of particles $N_b$ obtained by the three-level approximation method and the exact value. As the coupling strength $g$ decreases, the result obtained by the three-level approximation method is closer and closer to the exact solution by solving the master equation. Here, we use dimensionless parameters: $\lambda_a=0.01$, $\gamma_a=10$, $\gamma_b=1$, and $\kappa_e=0$ . }
\end{figure}
We consider a weak coupling $g$, $g\ll\gamma_a/\lambda_a$. The effective two-photon driving is weak. Therefore, the subsystem $b$ is at a low energy level. We consider the three-level subsystem $b$ with $\rho_{11}=1-\rho_{00}-\rho_{22}$. Keeping only the lowest three energy levels, Eq.~(\ref{eq:30}) is reduced to the following formulas
\begin{align}
\dot{\rho}_{00}=&-i\sqrt{2}\lambda_a g/\gamma_a[\rho_{20}-\rho_{02}]+2\gamma_b\rho_{11}+4\kappa'\rho_{22},\\
\dot{\rho}_{10}=&\gamma_b[2\sqrt{2}\rho_{21}-\rho_{10}],\nonumber\\
\dot{\rho}_{22}=&-4(\kappa'+\gamma_b)\rho_{22}-i\sqrt{2}\lambda_a g/\gamma_a(\rho_{02}-\rho_{20}),\\
\dot{\rho}_{21}=&-(2\kappa'+3\gamma_b)\rho_{21}-i\sqrt{2}\lambda_a g/\gamma_a(\rho_{01}),\\
\dot{\rho}_{20}=&-(2\kappa'+2\gamma_b)\rho_{20}-i\sqrt{2}\lambda_a g/\gamma_a(\rho_{00}\rho_{22}),
\end{align}
where $\kappa'=\kappa+\kappa_e$.
Let the left terms of the above equations be 0, the steady-state solutions are given by
\begin{align}
\rho_{00}=\frac{2(2g^2+\gamma_a(\kappa_e+\gamma_b))^2+g^2\lambda_a^2}{2(2g^2+\gamma_a(\kappa_e+\gamma_b))^2+4g^2\lambda_a^2},\\
\rho_{11}=\frac{g^2\lambda_a^2}{(2g^2+\gamma_a(\kappa_e+\gamma_b))^2+2g^2\lambda_a^2},\\
\rho_{22}=\frac{g^2\lambda_a^2}{2(2g^2+\gamma_a(\kappa_e+\gamma_b))^2+4g^2\lambda_a^2},\\
\rho_{20}=\frac{-ig(2g^2+\gamma_a(\kappa_e+\gamma_b))\lambda_a}{2(2g^2+\gamma_a(\kappa_e+\gamma_b))^2+4g^2\lambda_a^2},\\
\rho_{10}=0, \rho_{21}=0.
\end{align}
By calculating the expected value $N_b=\langle b^\dagger b\rangle$ from the three-level approximation method  and the master equation in Eq.~(\ref{eq:12}), we can find that the three-level approximation is more and more accurate as the coupling strength $g$ decreases, as shown in Fig.~\ref{fig.2}.

With the direct photon detection, the measurement uncertainty of the coupling strength $g$ is given by
\begin{align}
\delta^2 g|_{g=0}=\frac{3\gamma_a(\kappa_e+\gamma_b)^2}{16\lambda^2}.
\label{eq:40}
\end{align}

With the homodyne detection,  the measurement uncertainty is given by
\begin{align}
\delta^2 g|_{g=0}=\frac{\gamma_a(\kappa_e+\gamma_b)^2}{\lambda^2}.
\label{eq:41}
\end{align}

When the density matrix is expressed in diagonal form with the eigenstate $|k\rangle$, $\rho=\sum_{k=0,1,2}E_k|k\rangle\langle k|$, the quantum Fisher information can be calculated by
\begin{align}
\mathcal{F}(g)=&\sum_{k,E_k>0}\frac{(\partial_g E_k)^2}{E_k}\nonumber\\
&+\sum_{k,k',E_k+E_{k'}>0}\frac{2(E_k-E_{k'})^2}{E_k+E_{k'}}\langle k|\partial_g |k'\rangle.
\end{align}
After a simple calculation according to the above equation, the quantum Fisher information is given by
\begin{align}
\mathcal{F}(g=0)=\frac{6\lambda^2}{\gamma_a(\kappa_e+\gamma_b)^2}
\end{align}

According to the quantum Cram\'{e}r-rao bound, the measurement uncertainty is given by
\begin{align}
\delta^2 g|_{g=0}\geq1/\mathcal{F}(g=0)=\frac{\gamma_a(\kappa_e+\gamma_b)^2}{6\lambda^2}.
\end{align}
Comparing the above equation with Eq.~(\ref{eq:40}) and Eq.~(\ref{eq:41}), it shows that the direct homodyne detection is close to the optimal measurement.

When there are single photon dissipation, i.e., $\gamma_b\neq0$, the characteristic time to steady state is given by $\tau=1/\gamma_b$ when the coupling strength $g\rightarrow 0$.
In the steady state, the total photon number is given by $N_b=\langle b^\dagger b\rangle=\frac{2g^2\lambda_a^2}{(2g^2+\gamma_a(\kappa_e+\gamma_b))^2+2g^2\lambda_a^2}$. When $g\rightarrow 0$, the total photon number $N_b=0$. This result is interesting. Although the total photon number in the steady state is close to 0 and the interaction time is finite, the measurement uncertainty of the coupling strength $g$ is not infinite. In other words, when the signal mode suffers from the single-photon and two-photon dissipation, the information of the weak coupling strength $g$ can still be obtained in the steady state.

\subsection{The adiabatic elimination condition is not satisfied}
In this subsection, we consider that the adiabatic elimination condition $\gamma_a\gg\gamma_b$ is not satisfied.

By taking the expected value of both sides of Eq.~(\ref{eq:8}) and Eq.~(\ref{eq:9}), we obtain that
\begin{align}
\dot{\langle a\rangle}&=-ig \langle b^2 \rangle-\gamma_a \langle a\rangle+\lambda_a,\\
\dot{\langle b\rangle}&=-2ig \langle a b^\dagger\rangle-\gamma_b \langle b\rangle.
\end{align}

Let $\dot{\langle a\rangle}=0$ and $\dot{\langle b\rangle}=0$, the equations of motion for the steady-state mean values of the operators are derived as
\begin{align}
0&=-ig \langle b \rangle_s^2-\gamma_a \langle a\rangle_s+\lambda_a,\\
0&=-2ig \langle a\rangle_s\langle b\rangle^*-\gamma_b \langle b\rangle_s,
\end{align}
where we use  a mean-field like approximation $\langle b^2 \rangle_s=\langle b \rangle_s^2$ and $\langle a b^\dagger \rangle_s=\langle a \rangle_s\langle b \rangle_s^*$.
We set that $\langle a \rangle_s=x_a+i y_a$ and $\langle b \rangle_s=x_b+i y_b$, where $x_a$, $x_b$, $y_a$ and $y_b$ are real numbers. The above equations can be rewritten
as
\begin{align}
0&=-\gamma_ax_a+2gx_by_b+\lambda_a,\\
0&=g(x_b^2-y_b^2)+\gamma_ay_a,\\
0&=\gamma_bx_b+2gx_ay_b-2gx_by_a,\\
0&=\gamma_by_b+2g(x_ax_b+y_ay_b).
\end{align}
By solving the above equations, we obtain
\begin{align}
\textmd{solution (1)}:&x_a=\frac{\lambda_a}{\gamma_a}, x_b=y_a=y_b=0;\\
\textmd{solution (2)}:&x_a=\frac{\gamma_b}{2g}, x_b=\pm\frac{\sqrt{2g\lambda_a-\gamma_a\gamma_b}}{2g},\nonumber\\
&y_a=0, y_b=\mp\frac{\sqrt{2g\lambda_a-\gamma_a\gamma_b}}{2g}.
\end{align}

When $2g\lambda_a-\gamma_a\gamma_b<0$, the solution (2) does not exist due to the fact that $x_b$ can not be a complex number.

By using that $a=\langle a\rangle_s+\delta a$ and $b=\langle b\rangle_s+\delta b$, we can obtain the linearized quantum Langevin equations by keeping the terms up to the first
order of quantum fluctuation: $\mathbf{h}=(\delta a, \delta a^\dagger, \delta b, \delta b^\dagger)^\top$
\begin{align}
\mathbf{\dot{h}}=\mathbf{W} \mathbf{h}+\mathbf{h}_{\textmd{in}}
\end{align}
where the evolution matrix $\mathbf{W}$ is described as
\[
\mathbf{W}=\left(
\begin{array}{ll}
-\gamma_a\ \ \ \ \ \ \ 0\ \ \ \ -2ig\langle b\rangle_s\ \ \ \ \ \ \ 0\\
\ \ 0\ \ \ \ \ \ -\gamma_a\ \ \ \ \ \ \ 0\ \ \ \ \ \ \ \ \ \ \ 2ig\langle b\rangle_s^*\\
-2ig\langle b\rangle_s^*\ \ \ 0\ \ \ \ -\gamma_b\ \ \ \ \ \ \ \ -2ig\langle a\rangle_s\\
\ \ 0\ \ \ \ \ \ \ 2ig\langle b\rangle_s\ \ \ 2ig\langle a\rangle_s^*\ \ \ -\gamma_b
  \end{array}
\right ),
\]
and the quantum noise operators are $\mathbf{h}_{\textmd{in}}=(\sqrt{2\gamma_a}a_{\textmd{in}}, \sqrt{2\gamma_a}a^\dagger_{\textmd{in}}, \sqrt{2\gamma_b}b_{\textmd{in}}, \sqrt{2\gamma_b}b_{\textmd{in}})^\top$.

When $2g\lambda_a-\gamma_a\gamma_b<0$, the steady-state solution ($\langle a\rangle_s=\lambda_a/\gamma_a $, $\langle b\rangle_s=0$) ensures that the system is stable.
It is due to the fact that  all eigenvalues of the evolution matrix $\mathbf{W}$ have negative real parts.

When $2g\lambda_a-\gamma_a\gamma_b>0$, the solution ($\langle a\rangle_s=\lambda_a/\gamma_a $, $\langle b\rangle_s=0$) makes the system unstable. In this case, ($\langle a\rangle_s= \frac{\gamma_b}{2g}$, $\langle b\rangle_s=\mp\frac{\sqrt{2g\lambda_a-\gamma_a\gamma_b}}{2g}(1-i)$) can make the system stable.
When $\gamma_b=0$, we achieve that $\langle b\rangle_s=\mp\sqrt{\lambda_a/{2g}}(1-i)$. It recovers the previous result as shown in Eq.~(\ref{eq:15}).

Generally, it is defined as the normal phase when$\langle b\rangle_s=0$; and it is defined as the superradiance phase when$\langle b\rangle_s\neq0$\cite{lab32}. As the driving strength $\lambda_a$ increases, the system changes from the normal phase to the superradiant phase.
The critical point of the phase transition occurs when the driving strength $\lambda_a$ is equal to the critical driving strength $\lambda_c=\frac{\gamma_a\gamma_b}{2g}$.

In the normal phase, we can obtain the analytical solution.
Supposing that the system has reached the steady state after a long-time evolution, the solution of the quantum fluctuation operator $\delta b$ is
\begin{align}
\delta b=\int_0^\infty dt&[e^{-\gamma_b t}\cosh(gt\lambda_a/\gamma_a)b_{\textmd{in}}(t)\nonumber\\
& -ie^{-\gamma_b t}\sinh(gt\lambda_a/\gamma_a)b^\dagger_{\textmd{in}}(t)].
\end{align}
Using the correlation functions of the noise operator $b_{\textmd{in}}(t)$, we can obtain the expected values
\begin{align}
\langle \delta b\rangle&=\langle \delta b^\dagger\rangle=0,\\
\langle\delta b^\dagger \delta b \rangle&=\frac{2g^2 \lambda_a^2}{\gamma_a\gamma_b-4g^2\lambda_a^2},\\
\langle(\delta b)^2 \rangle&=\frac{-ig \lambda_a\gamma_a\gamma_b}{2(\gamma_a\gamma_b-4g^2\lambda_a^2)}.
\end{align}
Using the decoupling relation
$\langle ABCD\rangle=\langle AB\rangle\langle CD\rangle+\langle AC\rangle\langle BD\rangle+\langle AD\rangle\langle BC\rangle-2\langle A\rangle\langle B\rangle
\langle C\rangle\langle D\rangle$
as shown in ref.\cite{lab33}, we obtain the expected value
\begin{align}
\langle(\delta b^\dagger \delta b)^2 \rangle&\approx2\langle(\delta b^\dagger \delta b) \rangle^2 +\langle\delta  b^\dagger \delta b \rangle+|\langle(\delta b)^2\rangle|^2-2|\langle \delta  b\rangle|^4\nonumber\\
&=\frac{3g^2\lambda_a^2\gamma_a^2\gamma_b^2}{(\gamma_a^2\gamma_b^2-4g^2\lambda_a^2)^2}.
\end{align}

With the direct photon detection $b^\dagger b$, the measurement uncertainty of $g$ is given by
\begin{align}
\delta^2 g&=\frac{\langle(\delta b^\dagger \delta b)^2\rangle -\langle(\delta b^\dagger \delta b)\rangle^2 }{|\partial_g\langle(\delta b^\dagger \delta b)\rangle |^2}\\
&=\frac{(\gamma_a^2\gamma_b^2-4g^2\lambda_a^2)^2(3\gamma_a^2\gamma_b^2-4g^2\lambda_a^2)}{16\lambda_a^2\gamma_a^4\gamma_b^4}.
\end{align}

We can find that the measurement uncertainty $\delta^2 g$ is close to 0 as the phase transition point approaches, i.e., $\lambda_a\rightarrow\lambda_c$.
As a cost, the time resources used will also tend to be infinite.

\subsubsection{Thermal noise with temperature $T$}
When the system is subjected to a dissipative environment with a non-zero temperature $T$, the correlated values of the noises of the subsystem $b$ are rewritten as
\begin{align}
\langle b^\dagger_{\textmd{in}}(t) \rangle&=\langle b^\dagger_{\textmd{in}}(t) \rangle=0,\\
\langle b^\dagger_{\textmd{in}}(t)b_{\textmd{in}}(t') \rangle&=n(T)\delta(t-t'),\\
\langle b_{\textmd{in}}(t') b^\dagger_{\textmd{in}}(t)\rangle&=[n(T)+1]\delta(t-t'),
\end{align}
where the average thermal photon number $n(T)=\frac{1}{\exp (\omega_2 T)-1}$.
With the direct photon detection $b^\dagger b$, the measurement uncertainty of $g$ is given by
\begin{align}
\delta^2 g&=\frac{\langle(\delta b^\dagger \delta b)^2\rangle -\langle(\delta b^\dagger \delta b)\rangle^2 }{|\partial_g\langle(\delta b^\dagger \delta b)\rangle |^2}\\
&=\frac{(\gamma_a^2\gamma_b^2-4g^2\lambda_a^2)^2[(3+2n)\gamma_a^2\gamma_b^2+4g^2\lambda_a^2(2n-1)]}{16(1+2n)\lambda_a^2\gamma_a^4\gamma_b^4}.
\end{align}
From the above equation, we can obviously see that the measurement uncertainty $\delta^2 g$ is also close to 0 as the phase transition point approaches. Near the critical point of the phase transition, (i.e. $\lambda_a\simeq\lambda_c$), we get a temperature-independent result
\begin{align}
\delta^2 g=\frac{(\gamma_a\gamma_b-2g\lambda_a)^2}{4\lambda_a^2\gamma a \gamma_b}.
\end{align}
It shows that the influence of the temperature of the noise bath on the measurement precision of the coupling parameters tends to 0 near the phase transition point. That is, it is robust to the thermal fluctuations from the dissipative environment.

\section{Quantum sensor of the driving strength $\lambda_a$}
In addition to measuring the coupling strength $g$, it can also be used to measure the driving strength $\lambda_a$.
By the similar calculation, in the case of $\gamma_b=0$, the measurement precision of $\lambda_a$ is
\begin{align}
\delta^2\lambda_a=\frac{\lambda_a (2g^2+\gamma_a \kappa_e)}{2g}.\label{eq:69}
\end{align}
This result shows that the measurement uncertainty of the driving strength $\lambda_a$ decreases with the decrease of $\lambda_a$. It means that the dissipative degenerate down-conversion system can be a sophisticated quantum sensor about the weak driving strength when the signal mode does not suffer from the single-photon dissipation.
With photon number of the signal mode $N_b=\frac{2g\lambda_a}{2g^2+\gamma_a \kappa_e}$,  the measurement precision of $\lambda_a$ is rewritten as
\begin{align}
\delta^2\lambda_a=\frac{\lambda_a^2 }{N_b}.\label{eq:70}
\end{align}
This means that for $\lambda_a$ only the quantum limit is obtained, and nothing like $g$ can obtain the super-Heisenberg limit in Eq.~(\ref{eq:15}). It is due to the fact that the generator of $g$ is nonlinear and the generator of $\lambda_a$ is linear.
Moreover, we find that the measurement precision of $\lambda_a$  can be enhanced by reducing the coupling strength $g$ without the extra two-photon dissipation.
However, when $g<\sqrt{\gamma_a\kappa_e}$, the measurement precision of $\lambda_a$  will decrease as $g$ decreases. For $g=\sqrt{\gamma_a\kappa_e}$, we can achieve the optimal precision of $\lambda_a$
\begin{align}
\delta^2\lambda_a^o=\lambda_a\sqrt{2\gamma_a \kappa_e }.\label{eq:71}
\end{align}

\section{Conclusion}
We have investigated the quantum metrology in the degenerate down-conversion system.
Due to the nonlinear generator of the coupling strength, in the closed system the super-Heisenberg limit can be obtained with the semi-classical initial state. And the measurement precision increases with the evolution time, which gives an intuitive explanation of the results obtained in the dissipation case.

When $\gamma_a\gg\gamma_b$, we obtain the analytical result by adiabatically eliminating the pump mode. In the case of $\gamma_b=0$, the direct photon detection is proved to the optimal measurement. The result shows that the measurement uncertainty $\delta^2g$
is close to 0 as $g\rightarrow 0$. As a trade-off, the time to steady state also tends to infinity as $g\rightarrow 0$. Unexpectedly, the above result can not be influenced when the signal mode suffers from two-photon dissipation. The intuitive explanation is that there are subspaces that are not subject to two-photon dissipation. In the case of $\gamma_b\neq0$, an interesting result shows that although the total photon number in the steady state is close to 0 and the interaction time is finite, the measurement uncertainty of the coupling $g$ is not infinite. It means that the information of the coupling strength $g$ can still be obtained in the steady state no matter how small $g$ is.

When the adiabatic elimination condition is not satisfied, there is a critical phase transition point between the normal phase and superradiance phase. We can find that the measurement uncertainty $\delta^2 g$ is close to 0 as the phase transition point approaches. As before, it takes an infinite amount of encoding time to get that $\delta^2 g=0$. And we show that the influence of the temperature of the
noise bath on the measurement precision of the coupling
parameters tends to 0 near the phase transition point.

Finally, the degenerate down-conversion system can be used as a quantum sensor to measure the coherent driving strength. We show that the measurement uncertainty of
the driving strength $\lambda_a$ decreases with the decrease of the driving strength. Only the quantum limit is obtained for $\lambda_a$, which is due to the fact that the generator of $g$ is nonlinear and the generator of $\lambda_a$ is linear. This inspires us to discuss how to convert linear generators into nonlinear generators for improving the measurement precision.
Moreover, we find that the measurement precision of $\lambda_a$  can be enhanced by reducing the coupling strength $g$ without the extra two-photon dissipation.
However, when $g<\sqrt{\gamma_a\kappa_e}$, the measurement precision of $\lambda_a$  will decrease as $g$ decreases.

It is worth further exploring the problems related to multi-photon absorption and multi-parameter simultaneous measurement in the down-conversion system. In experiment, the down-conversion system can be observed typically in uniaxial and noncentrosymmetric media\cite{lab34}, a linear trapped three-ion crystal\cite{lab35}, and a lithium niobate crystal\cite{lab36}.

 \textit{Acknowledgements.-}This research was supported by the National Natural Science Foundation of China (Grant No. 12365001 and No. 62001134), and Guangxi Natural Science Foundation ( Grant No. 2020GXNSFAA159047).

\end{document}